\documentclass[aps,prb,twocolumn]{revtex4}
\usepackage{graphicx}
\usepackage{epsfig}
\usepackage{amsmath}
\usepackage{graphics}
\usepackage{epsfig}
\usepackage{tikz}
\usetikzlibrary{decorations.pathmorphing}

\newcommand{\bra}[1]{\ensuremath{\langle{#1}|\,}}
\newcommand{\ket}[1]{\ensuremath{\,|{#1}\rangle}}

\begin{document}

\title{Waiting time between charging and discharging processes in molecular junctions}

\author{Daniel S. Kosov}
\address{College of Science and Engineering, James Cook University, Townsville, QLD, 4811, Australia 
}


\begin{abstract}
When electric current flows through a molecular junction, the molecule constantly charges and discharges by  tunnelling electrons. These charging and discharging events 
occur at specific but random times and separated by stochastic time intervals. 
These time intervals can be associated with  dwelling time for a charge (electron or hole) to reside on the molecule. 
In this paper, the statistical properties of these time intervals are studied and general formula for their distribution is derived. The theory is based on 
the Markovian master equation which takes into account  transitions between vibrational states of charged and neutral molecule in the junction.  Two  quantum jump operators  are identified from the Liouvillian of the master equation - one  corresponds  to charging of the molecule and the other discharges the molecule back to neutral state. The quantum jump operators define the conditional probability that given that the molecule was charged by a tunnelling electron at time $t$, the molecule becomes neutral at later time $t+\tau$ discharging the electron to the drain electrode.  Statistical properties of these time intervals $\tau$
are studied  with the use of this distribution.
\end{abstract}

\maketitle
\section{Introduction}
Single-molecule electronics is the active field of research in chemical physics which has recently seen many advances.\cite{thoss-evers-review,doi:10.1021/acs.chemrev.5b00680} The molecular electronics has been promoted for decades as something to replace silicon based electronics (the dreams that never really came through), but rather it has  successfully grown into an integral part of modern chemical physics that gives unique and otherwise unavailable opportunities to study the fundamental issues of quantum mechanics and non-equilibrium statistical physics  of individual electron transfer events.\cite{doi:10.1021/acs.chemrev.5b00680,nitzan03,galperin07,nichols2010}

 One of the main feature that distances molecular electronics from  other nanoscale electron transport systems is the structural "softness" and as a consequence the possibility 
 to observe 
 current-induced "chemistry".\cite{PhysRevLett.78.4410,PhysRevLett.84.1527,ho99,PhysRevLett.85.2777,Repp26052006,dzhioev11,PhysRevB.86.195419,fuse,catalysis12,peskin2018,doi:10.1021/ja512523r,darwish2016,Pobelov:2017aa,doi:10.1021/acs.jpclett.8b00940} 
The electric current is an average quantity - it tells us how much energy is dissipated per unit time in the molecule, but current-triggered reactions do not only depend on the power pumped into the molecule but also on  dwelling time allowed for an extra electrons (or holes) to reside on the molecular bridge (if the molecule remains charged for a considerable time when the electric field between electrodes has a chance to produce a significant deformation of the molecular geometry).

 Electric  current  is a series of single electron quantum  tunneling events separated by random time intervals, that means the molecular bridge undergoes the continuous sequence of charging and discharging events also separated by random time intervals. This paper focuses on  the statistical properties of these time intervals and address the following questions. 
How long does the fluctuating charge stays on the molecule when the current flows through it? What is the distribution of these times and what are the statistical properties   of this distribution? How do  the vibrational dynamics and the coupling between electronic and vibrational degrees of freedom influence the distribution of the charging times?

This work is based on the ideas of the waiting time distribution (WTD) -- the theoretical approach to study statistics of individual electron tunnelling events in nanoscale systems.\cite{brandes08,buttiker12}
WTD is an extension of widely used in quantum transport methods of  full counting statistics.\cite{PhysRevB.84.205450,PhysRevB.87.115407,PhysRevB.91.235413,avriller09,thoss14,segal15,esposito13,nazarov-fcs,nazarov-book,bagrets11,cohen18} WTD has  recently gained a significant popularity in quantum transport research due to its intuitively clear interpretation and flexibility in design of various extensions   for a wide range of statistical 
applications.\cite{wtd-transient,flindt13,flindt14,flindt15,flindt17,kosov17-wtd,PhysRevB.92.125435,sothmann14,harbola15,rudge16a,rudge16b,PhysRevB.95.045306,kosov17-nonren}
  Traditional WTD is a conditional probability distribution that we observe the electron transfer in the detector electrode  at time $t+\tau$ given that an electron was detected in the same electrode at time $t$.
WTDs  are measured experimentally using time-resolved charge detection techniques for single-electron tunnelling.\cite{wtd-exp_2009} The charge detection is usually implemented by monitoring changes in electric current in auxiliary quantum point contact capacitively coupled to the main system.\cite{Lu2003,doi:10.1063/1.1815041,wtd-exp_2009,Ubbelohde:2012aa,PhysRevLett.96.076605}  Most experiments have been conducted at ultra-low temperatures (from mK to several K), although  room temperature measurements were reported for a carbon nanotube by monitoring optical blinking of semiconductor nanocrystal which is induced by the charging/discharging events in the nanonotube.\cite{doi:10.1021/acs.nanolett.5b01338}
The main limitation of all  single-electron counting experimental methods is the restriction of counting no more than of approximately $10^3$ electrons per second (that is very low electric current).  The interesting hybrid approach was proposed to extract WTD directly from low-order correlation experimental measurements via theoretical post-processing using continuous matrix product state tomography.\cite{1367-2630-17-11-113024}

In our recent studies,\cite{kosov17-wtd, kosov17-nonren} we used WTD to analyse the electron transport through molecular junctions with electron-vibration coupling.
In this paper, WTD notion is adopted to study different kind of probability distribution which is  related to the temporal fluctuations of the molecular charging state rather than electric current, namely, we will define and explore the {\it  conditional probability distribution that   given that the molecule was charged by the tunneling electron at time $t$, the molecule becomes neutral at later time $t+\tau$ discharging an electron to the drain electrode}.  
This type of WTD follows the same philosophy as  tunneling and residence time distributions explored in papers.\cite{rudge16a,rudge16b}
This distribution will be  used in our paper to study statistical properties of temporal fluctuations of molecular charging states in a current-carrying junction with electron-vibration interaction.

The paper is organised as follows. Section II describes the master equations, defines charging and discharging quantum jump operators, and gives the derivation of main equations for WTDs. In section III, we  present analytical and numerical study of statistical properties of the waiting times between charging and discharging events in a molecular junction with electron-vibration interaction.  Section IV summarises the main results of the paper. 

We use natural  units for quantum transport  throughout the paper: $\hbar=k_{B}=e=1$.

\section{Theory}
To have a specific model, let us suppose that when an electron is transferred to the molecule from the electrodes, the molecule  becomes negatively charged and when this electron leaves, the molecule comes back to the neutral state. This scenario corresponds to the case of electron transport through a single resonant level above the equilibrium Fermi energy of the electrodes. The opposite case of a molecular bridge
acquiring  a positive charge (transport through a resonant level which is below the Fermi energy) can be considered likewise and we will simply give the final expression for this process for the comparison in the end of this section.

We begin with the following intuitively obvious master equation
\begin{eqnarray}
\dot P_{0q}(t) &=& \sum_{\alpha q'} \Gamma^\alpha_{0q,1q'} P_{1q'} (t)  -  \Gamma^\alpha_{1q', 0q} P_{0q}(t),
\label{me1}
\\
\dot P_{1q}(t)&=& \sum_{\alpha q'} \Gamma^\alpha_{1q,0q'} P_{0q'}(t)  -  \Gamma^\alpha_{0q',1q} P_{1q}(t),
\label{me2}
\end{eqnarray}
where  $P_{0q}(t)$ is the probability that the molecule is neutral  and occupied by $q$ vibrational quanta at time $t$,
 $P_{1q}(t)$ is the probability that the molecule is charged  and occupied by $q$ vibrational quanta at time $t$.
The transition rates and the model are explained in Figure 1. The rate 
$\Gamma^\alpha_{0q',1q} $
describes the  transition from charged state and $q$ vibrations to the electronically neutral state with $q'$ vibrations  by the electron transfer from the molecule to $\alpha=S,D$  electrode
and the rate $\Gamma^\alpha_{1q',0q} $ describes the transition  to charged state from the originally neutral molecule  by electron transfer from $\alpha$  electrode simultaneously changing the vibrational state from $q$ to $q'$.

 \begin{figure}[t]
\begin{center}
\includegraphics[width=1.0\columnwidth]{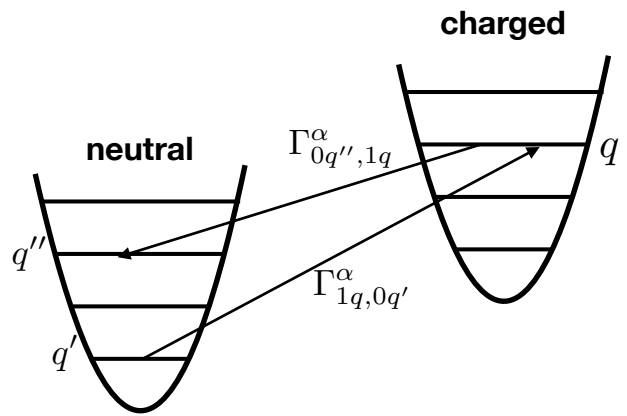}
\end{center}
	\caption{Sketch of the model. $\Gamma^\alpha_{mq,nq'}$ are the rates for the molecular charging or discharging processes by the transition of electron to/from  source $\alpha=S$ and drain $\alpha=D$ electrodes by changing the molecular electronic state from $n$ to $m$ and simultaneously changing the molecular vibrational state from $q'$ to $q$.}	
\label{sketch}
\end{figure}

We will work with the  probability vector ordered such  that the probabilities to observe the molecule in neutral and charged states are grouped in pairs of equal number of the vibrational quanta:  
\begin{eqnarray}
\label{P}
\mathbf P(t) =
\begin{bmatrix}
P_{00} (t)\\
P_{10}(t)\\
{P_{01}(t)}\\
{P_{11}(t)}\\
\vdots
\\
{P_{0N}(t)}\\
{P_{1N}(t)}\\
\end{bmatrix} ,
\end{eqnarray}
where $N$ is the total number of vibrational quanta  included into the calculations.
It is useful for our derivations to define the identity  vector of length $2N$:
\begin{eqnarray}
\mathbf I=
\begin{bmatrix}
{1}\\
{1}\\
{1}\\
{1}\\
\vdots
\\
{1}\\
{1}\\
\end{bmatrix} .
\end{eqnarray}
The normalisation of the probability is given by the scalar product 
\begin{equation}
 ( \mathbf I,  \mathbf P(t))= 1.
\end{equation}

Using $\mathbf P(t)$ we write the master equation (\ref{me1},\ref{me2})  in a matrix form
\begin{equation}
\dot {\mathbf P}(t) = {\cal L} \mathbf P(t),
\label{master-eq}
\end{equation}
where $\cal L$ is the Liouvillian operator.  From the Liouvillian operator we can identify two quantum jump operators corresponding to the processes of molecular charging and discharging by tunnelling electrons.
These quantum jump operators are $2N\times 2N$   matrices which are determined by considering their actions on the probability vector.
For our discussion of charging and discharging processes it is sufficient  to define the two jump operators:
the jump operator $J_d$ for transferring an electron from the molecule to the drain electrode (discharging in our case) 
\begin{equation}
({ J}_d \mathbf P(t) )_{nq} =  \delta_{n0} \sum_{q'} \Gamma^{D}_{0q,1q'} P_{1q'} (t),
\label{jumpD}
\end{equation}
and the operator $J_c$ which describes electron transfer
from the source electrode to the molecule (charging)
\begin{equation}
({ J}_c\mathbf P(t) )_{nq} =  \delta_{n1} \sum_{q'} \Gamma^{S}_{1q,0q'} P_{0q'} (t). 
\label{jumpC}
\end{equation}
Each quantum jump is  associated with the   corresponding quantum measurement operator.
If $\mathbf P(t)$ is the probability vector before the measurement, then after the quantum jump detection it collapses to the vector\cite{Petruccione}
\begin{equation}
M_{c/d} \mathbf P(t) = \frac{ J_{c/d} \mathbf P(t)}{(\mathbf I, J_{c/d} \mathbf P(t))},
\label{m}
\end{equation}
where $M_{c/d}$ is the quantum measurement operator related to the jump $J_{c/d}$.

We first extract from the total Liouvillian the part which generates the  evolution without electron transfer from the molecule to the drain electrode
\begin{equation}
{\cal L} _0= {\cal L}-J_d,
\end{equation}
and re-write master equation  (\ref{master-eq}) as
\begin{equation}
\dot {\mathbf P}(t) = ({\cal L}_0+ J_d) \mathbf P(t).
\label{master-eq2}
\end{equation}
Next,  (\ref{master-eq2}) is converted  to the integral equation
\begin{equation}
\mathbf P(t) =e^{{\cal L}_0 t}   \mathbf P(0) + \int_0^t dt_1 e^{{\cal L}_0 (t-t_1)}  J_d \mathbf P(t_1),
\label{master-int}
\end{equation}
which then is resolved as a series
\begin{align}
\label{master-int1}
\mathbf P(t) =e^{{\cal L}_0 t}   \mathbf P(0) 
+  \int_0^t d t_1  e^{{\cal L}_0  (t-t_1)}  J_d    e^{{\cal L}_0 t_1} \mathbf P(0)
 + ....
\nonumber
\end{align}
We truncate this series at the first integral and that is sufficient to define the required time distributions. The next order terms in this 
expansion can be considered to define higher order probability distributions which describe correlations between two and more waiting times, for detailed discussion we refer to appendix A in paper.\cite{kosov17-nonren}

The initial probability vector  is chosen as
\begin{equation}
\mathbf P(0) = M_c \mathbf P,
\end{equation}
where $\mathbf P$  describes the non-equilibrium steady state. This choice of the initial state implies  that the charging of the molecule by an electron transfer from the source electrode is detected at time $t=0$ in the steady state regime and then we start to monitor the system for  discharging events:
\begin{align}
\label{master-int2}
&\mathbf P(t) =e^{{\cal L}_0 t}   M_c \mathbf P+  \int_0^t d t_1  e^{{\cal L}_0 (t-t_1)}  J_d    e^{{\cal L}_0 t_1} M_c \mathbf P.
\end{align}
Using electron detection operator (\ref{m}),
we rewrite (\ref{master-int2}) in a form which explains the probabilistic meanings of the integral term and, consequently, enables 
extraction of an expression for the probability distribution of the  time delay between charging and 
discharging events:
\begin{multline}
\label{master-int3}
\mathbf P(t) =e^{{\cal L}_0 t}   M_c \mathbf P
\\
+  \int_0^t d t_1  \frac{(\mathbf I, J_d e^{{\cal L}_0 t_1} J_c \mathbf P)}{(\mathbf I, J_c \mathbf P)} \;  e^{{\cal L}_0 (t-t_1)}   M_d    e^{{\cal L}_0 t_1} M_c \mathbf P
\end{multline}
Let us now interpret (\ref{master-int3}). The first term, $ e^{{\cal L}_0 t}  M_c  \mathbf P$, is the contribution to the probability vector  from all measurements  where no electron transfer from the molecule to the drain electrode to occur up to time $t$ after the initial molecular charging at time $t=0$.  The physical  meaning of the integral term in (\ref{master-int3}) is deduced using the following arguments.\cite{Srinivas2010,Zoller1987,nonrenewal-budini,kosov17-nonren} The molecule is charged by  transferring electron into the molecule from the source electrode at time $t=0$ (term $M_c \mathbf P$), then no detection of an electron transfer to the drain electrode occurs up to time $t_1$ ("idle" evolution operator $e^{{\cal L}_0 t_1}$),
then the detection of the discharging of the molecule by transferring an electron to the drain electrode is observed at time $t_1$ (quantum measurement operator $M_d$),  and finally the system "idle" without electron transfer up to time $t$.
 Therefore, the pre-factor $(\mathbf I, J_d e^{{\cal L}_0 t_1} J_c \mathbf P)/(\mathbf I, J_c \mathbf P)$ should be understood as the probability of observing this process.

 The main result of this section is summarized in the equation below. The distribution of waiting times between charging and discharging events in molecular junction is                                     
\begin{equation}
 w(t)=\frac{(\mathbf I, J_d e^{{\cal L}_0 t} J_c \mathbf P)}{(\mathbf I, J_c \mathbf P)}.
 \label{w}
\end{equation}
The derivations for fluctuating positive charge (transport through highest occupied molecular orbital resonant level) can be performed along exactly the same lines and give
\begin{equation}
 w(t)=\frac{(\mathbf I, J_c e^{{\cal L}_0 t} J_d \mathbf P)}{(\mathbf I, J_d \mathbf P)}.
 \label{ww}
\end{equation}
These distributions are conceptually similar to tunnelling and residence time distributions proposed for electron transport through a single resonant level,\cite{rudge16a,rudge16b} though we use different definition of the "idle" Liouvillian here. It is instructive to contrast definitions (\ref{w}) and (\ref{ww}) with standard WTD used in electron transport theory which  is a conditional probability to observe the electron transfer in the detector electrode given that an electron was detected in the {\it same} detector  electrode at earlier time  ${(\mathbf I, J_d e^{{\cal L}_0 t} J_d \mathbf P)}/{(\mathbf I, J_d \mathbf P)}$.\cite{brandes08}

We would like to make an important note on the range of validity of the developed theory. The expressions for WTD derived here technically work for bidirectional and unidirectional electron transport. However, the rigorous interpretation of (\ref{w},\ref{ww}) as a conditional distributions of delay times between charging and discharging events is possible only in the absence of the electron back-tunnelling against the average current flow (tunnelling from the drain electrode to the molecule or from the molecule to the source electrode). If the back-tunnelling is not suppressed by the choice of the applied voltage and other parameters of the model, the charging events from the reverse tunnelling electrons become physically present but they are concealed inside the idle time-evolution operator $e^{{\cal L}_0 t} $ and are not being explicitly monitored in  (\ref{w},\ref{ww}). For the case of bidirectional current, one should think about WTDs given by (\ref{w},\ref{ww}) as conditional distribution of waiting time between events of electron tunnelling to the drain electrode and electron tunnelling from the source electrode (with back-tunnelling charging and discharging events covert in the waiting time intervals).

 \section{Model calculations} 
 Our study of statistical properties of time delays separating charging and discharging events will be based on the following idealised model. The molecule is attached to two macroscopic leads (source and drain) held at different chemical potentials and is  represented by a single molecular orbital linearly coupled to local  vibration.
  The total Hamiltonian of molecular junction is
\begin{equation}
H= H_{\text{molecule}} + H_{\text{electrodes}} + H_T.
\end{equation}

 The molecular Hamiltonian is
\begin{equation}
H_{\text{molecule}}= \epsilon_0 a^\dag a  +   \omega b^\dag b+ \lambda a^\dag a (b^\dag + b),
\end{equation}
where $\epsilon_0$ is the energy or the molecular orbital,  $\omega$ is the vibrational frequency, and $\lambda$ is the strength of  electron-vibration coupling. $a^\dagger (a) $  creates (annihilates) an electron on molecular orbital, and $b^+ (b)$ is bosonic creation (annihilation) operator for the molecular vibration.

Electrodes consist of noninteracting electrons:
\begin{eqnarray}
H_{\text{electrodes}}=  \sum_{k\alpha} \epsilon_{k\alpha} a^\dag_{k \alpha} a_{k \alpha},
\end{eqnarray}
where $a^\dagger_{k\alpha}$   creates  an electron in the single-particle state $k$ of $\alpha$ electrode and $a_{k\alpha}$ is the corresponding electron  annihilation operator. The electron tunnelling is described by 
\begin{eqnarray}
H_T=  \sum_{k\alpha }( t_{k\alpha}   a^\dag_{k \alpha} a  + h.c),
\end{eqnarray}
where $t_{k\alpha}$ is the tunnelling amplitudes.

Lang-Firsov  unitary rotation (polaron transformation) of molecular operators \cite{lang_firsov1963} is used to remove electron-vibration coupling from the molecular Hamiltonian: 
\begin{equation}
a= \tilde a e^{\nu(\tilde b^\dag -\tilde b)}, \;\;\;\; b = \tilde b + \nu \tilde a^\dag \tilde a,
\end{equation}
where $\tilde a^\dag (\tilde a )$ and $\tilde b^\dag (\tilde b)$ are  new  creation (annihilation) operators for molecular electron and vibration.
The molecular Hamiltonian becomes 
\begin{equation}
H_{\text{molecule}} = \epsilon \tilde a^\dag \tilde a + \omega \tilde b^\dag \tilde b,
\label{hm}
\end{equation}
where the molecular orbital energy $\epsilon$   includes polaron shift
$\epsilon = \epsilon_0 - \lambda^2/ \omega $.
The Hamiltonian for the electrodes  is invariant under Lang-Firsov rotation and the tunnelling interaction becomes
\begin{eqnarray}
H_T=   \sum_{k \alpha  } (t_{k\alpha}  e^{-\frac{\lambda}{\omega} (\tilde b^\dag -\tilde b)} a^\dag_{k\alpha} \widetilde a  + h.c).
\end{eqnarray}
Now, after Lang-Firsov transformation, the  molecular Hamiltonian (\ref{hm}) is diagonal.
Next,  the application of the sequence of approximations:  the Born approximation (keeping terms up to second order in $H_T$ in the Liouville equation for the reduced density matrix), the Markov approximation (assumption that the correlation functions of the electrodes decay on a time scale much faster than tunneling events) and the secular approximations (amounts to neglect coherences between charge states of the molecule) leads to the master equation (\ref{me1},\ref{me2}) with the following transition rates:\cite{PhysRevB.69.245302} 
\begin{equation}
\Gamma^\alpha_{0q',1q} =  \gamma^\alpha |X_{q'q}|^2 \left(1-f_\alpha[\epsilon-\omega (q'-q)] \right)
\end{equation}
and
\begin{equation} 
\Gamma^\alpha_{1q',0q} =    \gamma^\alpha |X_{q'q}|^2  f_\alpha[\epsilon+\omega (q'-q)].
\end{equation}
The rates depend on Fermi-Dirac occupation numbers for the electrode states
$f_\alpha$,
Franck-Condon factor $X_{qq'}$,
and electronic level broadening function  $\gamma^\alpha$.
The Franck-Condon factor 
\begin{equation}
X_{qq'}= \bra{q} e^{-\lambda/\omega (b^\dag  -b)} \ket{q'} 
\end{equation}
is determined by the strength of the electron-vibration coupling $\lambda$.

 Master equation (\ref{me1},\ref{me2})  describes  non-equilibrium dynamics of the molecular vibrations and 
this fully  non-equilibrium case will shortly be considered numerically. First,
we take the limit where the vibration is maintained in thermodynamic equilibrium at some temperature $T$; it enables us to obtain  analytical expressions for the probability distributions. This limit physically means that the molecular vibration is attached to its own bath, which can be, for example, a  solvent around the molecular junction or surface phonons in metal electrodes.

 To implement this limit we use the following separable ansatz for the probabilities\cite{PhysRevB.69.245302}
 \begin{equation}
 P_{nq}(t) =P_{n}(t) { \frac{e^{-q \omega/T}}{1-e^{-\omega/T}}},
 \end{equation}
 which assumes that  the vibration maintains the equilibrium distribution at all time.
The master equation (\ref{me1},\ref{me2}) is reduced to evolution equation for the probabilities to observe the molecule in neutral  and charged states, $P_0$ and  $P_1$, respectively:
\begin{eqnarray*}
\frac{d}{dt} \left[\begin{array}{c}
 P_0\\
 P_1
\end{array}\right] & = & \left[\begin{array}{cc}
- \Gamma_{10} &  \Gamma_{01}\\
 \Gamma_{10} & - \Gamma_{01}
\end{array}\right]\left[\begin{array}{c}
P_0\\
P_1
\end{array}\right],
\label{rate}
\end{eqnarray*}
where the vibration averaged  rates are defined as
\begin{equation}
 \Gamma^\alpha_{mn} = \sum_{qq'} \Gamma^\alpha_{mq,nq'} \frac{e^{-q' \omega/T}}{1-e^{-\omega/T}},
\end{equation}
and the total rates include contributions from the source and drain electrodes $ \Gamma_{mn}= \sum_\alpha  \Gamma^\alpha_{mn}$.
We write discharging and charging quantum  jump operators in a matrix form and  also as a  dyadic product of two vectors
\begin{equation}
 {J_d}  = \left[\begin{array}{cc}
0 &  \Gamma_{01}^D \\
0 & 0
\end{array}\right]= 
\Gamma^D_{01}
 \left[\begin{array}{c}
1\\
0
\end{array}\right]\left[\begin{array}{cc}
0 &  1  \end{array}\right],
\end{equation}
\begin{equation}
 {J_c}  = \left[\begin{array}{cc}
0 &  0 \\
\Gamma_{10}^S & 0
\end{array}\right]= 
\Gamma^S_{10}
 \left[\begin{array}{c}
0\\
1
\end{array}\right]\left[\begin{array}{cc}
1 &  0  \end{array}\right].
\end{equation}
Next, straightforward vector algebra brings the WTD between charging and discharging events to 
 \begin{equation}
w_{eq}(\tau) = \Gamma_{01}^D  
\left[\begin{array}{cc}
0 & 1 \end{array}\right]
e^{ {\cal L}_0 \tau }
\left[\begin{array}{c}
0\\
1
\end{array}\right].
\end{equation} 
where
\begin{equation}
 {\cal L}_0  = {\cal L} -J_d= \left[\begin{array}{cc}
- \Gamma_{10} &  \Gamma^S_{01}\\
 \Gamma_{10} & - \Gamma_{01}
\end{array}\right].
\end{equation}
Here  "eq"  subscript in the time distribution  indicates the equilibrium molecular vibration (and, of course,  electrons are still in non-equilibrium).
 This expression can be further evaluated and brought to the following analytic form
 \begin{multline}
w_{eq}(\tau)=
\frac{ \Gamma_{01}^D}{2Z}\Big\{ (Z+\Gamma_{01}-\Gamma_{10})   e^{-(Z+\Gamma_{01}+\Gamma_{10}) t/2} \\
+   (Z+\Gamma_{10}-\Gamma_{01}) e^{-(\Gamma_{01}+\Gamma_{10}-Z) t/2}
\Big\},
\label{weq}
  \end{multline}
 where
 \begin{equation}
 Z= \sqrt{(\Gamma_{01}-\Gamma_{10})^2+4 \Gamma_{10}\Gamma^S_{01}}.
 \end{equation}
 The bi-exponential dependence of the distribution function is a consequence of rare extreme events, electron tunnelling  against the current flow from the molecule to the source electrode or from the drain electrode back to the molecule. Suppressing the back-scattering electron transfer,
\begin{equation}
\Gamma^D_{10} \rightarrow 0, \;\;\;  \Gamma^S_{01} \rightarrow 0, 
\end{equation}
yields  single-exponent WTD
      \begin{equation}
w_{eq}(\tau)=
\Gamma_{01}^D   e^{-\Gamma^D_{01} \tau}. 
\label{w_single}
\end{equation}

    \begin{figure}[t]
\begin{center}
\includegraphics[width=1.1\columnwidth]{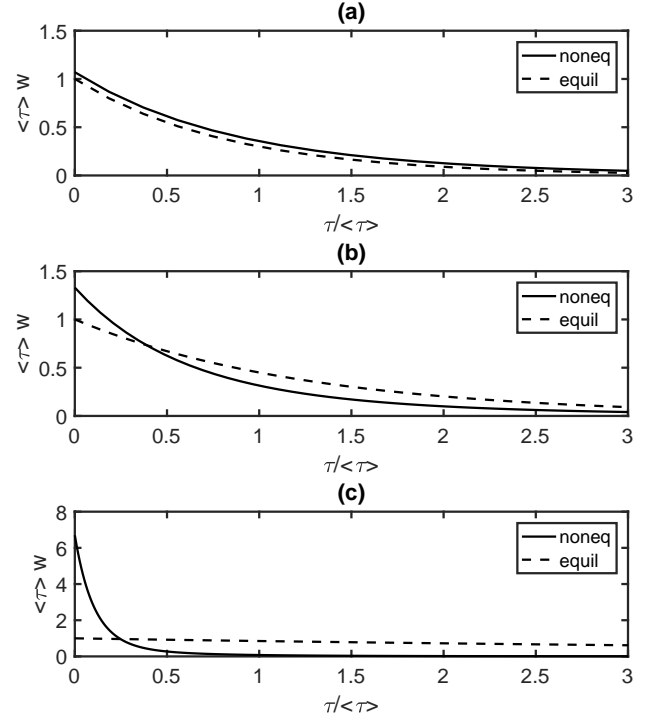}
\end{center}
	\caption{WTD between molecular charging and discharging events  computed for  different values of the electron-vibrational coupling strength: (a) $\lambda=1$, (b) $\lambda=2$, and (c) $\lambda=3$. Parameters used in calculations: $\omega=1$, $\gamma_S = \gamma_D =0.01$, $ T =0.05$, 
$\epsilon =0.1$, $V_{sd}=4$.  All energy values are given in units of $\omega$.}	
\label{fig2}
\end{figure}

 Fig. 2 shows the WTDs for between molecular charging and discharging processes computed for different values of the electron-vibration coupling strength.  WTDs for equilibrated  vibration are calculated using (\ref{weq}) and fully non-equilibrium WTDs are
 computed numerically via (\ref{w}).  All distributions attain their maximum values at $\tau=0$, therefore  
 the mode waiting times between charging and discharging process is always zero. 
 It indicates that the short time behaviour of time delays between charging and discharging events is dominated by the Poisson point process noise.
It is not clear at the moment if this short time behaviour is merely a result of  approximations used to derive the underlying master equation (Markov and secular approximations, in particular).
 The electron-vibrational interaction has the opposite effect on the  behaviour of rime distributions for equilibrium and non-equilibrium  molecular vibrations. The increasing strength of electron-vibrational interaction  significantly squeezes the non-equilibrium time distribution but   the equilibrium one  is just sightly stretched to the longer than average waiting times. These different behaviours can be understood based on the following  considerations. In the equilibrium regime, the molecular vibrational state is forced to be  populated to a given  temperature irrespective to electronic  degrees of freedom and, as a result, the time distribution function (when scaled by its average waiting time $\langle \tau \rangle$) shows  only small dependence on the strength of the electron-vibration coupling. That means that in the case of forcefully equilibrated molecular vibrations, the strength of the electron-vibration coupling affect mostly  the average value leaving the other WTD parameters  intact.
 Contrary, in non-equilibrium case, the different numbers of vibrational quanta can  be dynamically excited and de-excited by tunneling electrons and, therefore, the electron-vibration interaction plays a critical role for these processes.

Figure 3 shows the average time delay between molecular charging and discharging events as a function of the voltage bias for different strengths  of the electron-vibration coupling.
The steps  in the average time is related to the resonant excitations of the vibration states by electric current which occur when the voltage passes through an integer multiple  of the vibrational energy.  The temperature effects smoothes the edges of these steps. This behaviour reflects the staircase dependence of the electric current on voltage;\cite{PhysRevB.69.245302} the sharp increase of charging state lifetime in the regime of strong coupling and low voltage  corresponds to the Franck-Condon blockade suppression of the electric current.\cite{fcblockade05}

Figures 4  shows the  relative standard deviation (RSD)
\begin{equation}
\text{RSD}= \frac{\sqrt{ \langle \tau^2 \rangle - \langle \tau \rangle^2 }}{\langle \tau \rangle}
\label{rsd}
\end{equation}
computed  for different strengths of electron-vibrational coupling as a function of  applied voltage bias. Notice that mathematically a single exponential arbitrary distribution has always RSD=1, irrespective to the parameters. At small voltages the RSDs are greater than 1 for both equilibrium and non-equilibrium vibrational dynamics indicating a multi-exponential character of the WTDs -- this is due to the  admixture of  back-scattering electron transfers from the molecule back to the source electrode or from the drain electrode to the molecule. When the backscattering processes are suppressed by the voltage bias ($V_{sd} \gtrsim 0.5$), the RSD becomes exactly 1 for the equilibrium vibrations  that means that $w_{eq}(\tau)$ is reduced to a single exponential form (\ref{w_single}). The WTD for non-equilibrium vibrations also becomes single exponential in the voltage range $0.5 \lesssim V_{sd} \lesssim 1.5$,  since here the backscattering processes are already suppressed by the voltage bias but  inelastic vibrational channels for electron transport have not been opened yet. The further increase of voltage above the excitation threshold for inelastic channels leads to the  increase  RSD, which means that the non-equilibrium WTD becomes a multi-exponential (i.e., multichannel) distribution.

The numerical calculations elucidate the origin of small spikes in the RSD voltage dependence seen in Figure 4 for the non-equilibrium vibrations. These spikes occur only at the opening of  new vibrational transport channels and they are due to electron backscattering events governed by the rates $\Gamma^D_{1q,0q'}$ and $\Gamma^S_{0q,1q'}$  with $q,q'>0$. If these rates are set to zero, the RSD spikes vanish.
The parallel can be drawn with the well-known phenomena in electron current noise - the noise is small when the single channel dominates the transport and grows once the electron transport is distributed along several channels.

 \begin{figure}[t]
\begin{center}
\includegraphics[width=1.0\columnwidth]{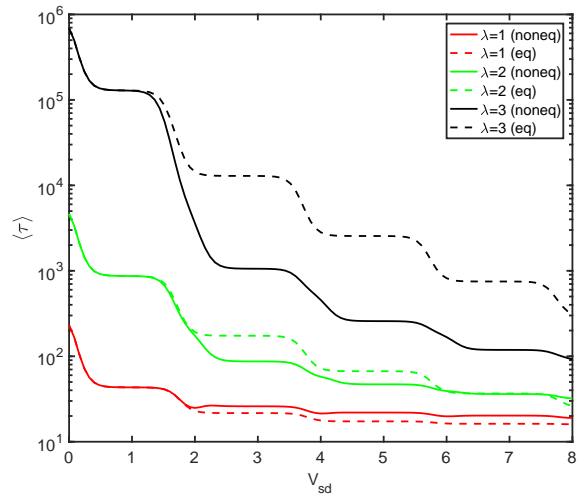}
\end{center}
	\caption{Average waiting time between charging and discharging processes  as a function of applied voltage  $V_{sd}$ computed for different values of electron-vibrational coupling $\lambda$. Parameters used in calculations (all energy values are given in units of $\omega$): $\omega=1$, $\gamma_S = \gamma_D =0.01$, $ T =0.05$, 
$\epsilon =0.1$.  The voltage bias $V_{sd}$ is given in $\omega$ and time is measured in  periods of the molecular vibration $2 \pi/\omega$.}	
\label{fig3}
\end{figure}

 \begin{figure}[t]
\begin{center}
\includegraphics[width=1.1\columnwidth]{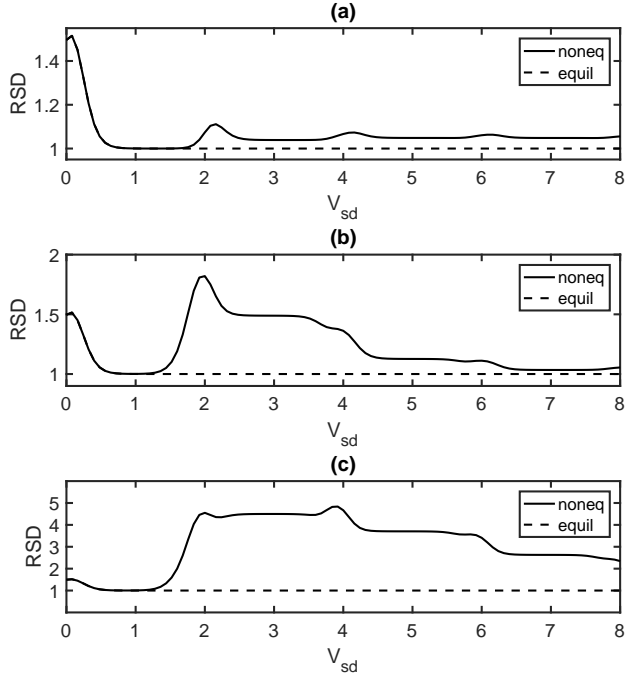}
\end{center}
	\caption{Relative standard deviation (\ref{rsd}) as a function of applied voltage  $V_{sd}$ computed for  different values of the electron-vibrational coupling strength: (a) $\lambda=1$, (b) $\lambda=2$, and (c) $\lambda=3$. Parameters used in calculations (all energy values are given in units of $\omega$): $\omega=1$, $\gamma_S = \gamma_D =0.01$, $ T =0.05$, 
$\epsilon =0.1$.  The voltage bias $V_{sd}$ is given in $\omega$.}	
\label{fig4}
\end{figure}

\section{Conclusions}
We have developed a theoretical approach to compute statistical distributions of waiting times between charging and discharging process in a molecular junction with electron-vibration interaction.
The approach is based on the use of Markovian master equation with exact treatment of electron-vibration interaction to describe electron transport through a molecular junction. Two quantum jump operators responsible for molecular bridge charging and discharging were extracted from the Liouvillian of the master equation.
These  jump operators were used to develop WTD for time delays between charging and discharging events. The statistics of these events were studied analytically and numerically for a model molecular junction 
described by the Holstein Hamiltonian.

The main observations are
\begin{itemize}
\item
For the case of equilibrium vibrations, the distribution of waiting times between molecular charging and discharging processes is bi-exponential -- the dominant exponent represent the electron transport from the molecule to the drain electrode, while the other exponent reflects the presence of the rare (at high voltage) electron tunneling events where an electron is moving against the average electric current flow.

\item
The WTDs between charging and discharging events have distinct dependence of the strength of electron-vibration coupling. For weak electron-vibration coupling ($\lambda=1$) there is not much difference between 
WTDs for equilibrium and non-equilibrium vibrations, whereas as the strength of electron-vibration coupling grows ($\lambda=2,3$), the non-equilibrium WTD narrows down to its mode time ($\tau=0$) but equilibrium WTD spreads to the much longer than its average waiting times. Once scaled by their  average waiting times,  the equilibrium WTD shows much less dependence on $\lambda$ than its non-equilibrium counterpart.

\item
Analysis of the WTDs using RSD (noise-to-signal ratio for measuring average waiting time between charging and discharging processes) 
shows that dynamical openings of elastic and inelastic vibrational transport channels   by the increasing voltage bias lead to significant increase of the RSD ($>1$)  for non-equilibrium vibrational dynamics indicating multi-exponential nature of non-equilibrium WTD.  
\end{itemize}

\begin{acknowledgments}
The author thanks Samuel Rudge for many valuable and stimulating discussions.
\end{acknowledgments}

\clearpage

\end{document}